\documentstyle[12pt]{article}
\newcommand{\beq}[1]{\begin{equation}\label{#1}}
\newcommand{\eeq}{\end{equation}}
\def\pipi{\mbox{$\pi^+\pi^-$}}

\begin{document}

\begin{center}
{\large\bf
Ratio of productions of \pipi{}-atoms to free \pipi{} pairs with account of
the strong interaction in final states\\}
\vspace*{0.5cm}
L.AFANASYEV and O.VOSKRESENSKAYA\\
{\it Joint Institute for Nuclear Research, \\
141980 Dubna, Moscow Region,\\
Russia}
\end{center}

\begin{abstract}
{\small
The ratio of productions of \pipi{}-atoms to free \pipi{} pairs is
calculated with account of the strong interaction in final states. It is
shown that this ratio is expressed via a squared ratio of the well-known
Coulomb wave functions and thus can be calculated with a high accuracy.
}

\end{abstract}

A method of measurement of the \pipi{}-atom lifetime proposed by L.L.Nemenov
\cite{nem85} and which is under realization in the experiment DIRAC
\cite{dirac} at CERN Proton Synchrotron is essentially based on the
assumption that the ratio $R(n\mathrm{s}/p\mathrm{s})$ of the \pipi{}-atom
production rate in $n$s states in hadron-nuclear interactions at high energy
to the production rate of so called "Coulomb" \pipi{} pairs (i.e. free
\pipi{} pairs with the relative momentum less or of order of the Bohr
momentum $\alpha m_\pi /2$ of the \pipi{}-atom) can be calculated with
accuracy better than 1\%.  Here we consider the accuracy of this ratio with
account of results of last papers \cite{af97,kur98}.

According to \cite{nem85,af97} the ration can be written as:
\beq{e1}
R(n\mathrm{s}/p\mathrm{s})=\frac{\displaystyle
\left|\int M(\vec r) \psi_{n\mathrm{s}}(\vec r)\,d^3r \right|^2 }
{\displaystyle
\left|\int M(\vec r) \psi_{p\mathrm{s}}(\vec r)\,d^3r \right|^2 } \, ,
\eeq
where $\psi_{n\mathrm{s}}(\vec r)$ is the \pipi-atom wave function with the
principal quantum number $n$ and zero orbital one;
$\psi_{p\mathrm{s}}(\vec r)$ is the wave function of \pipi{} pair with
the relative momentum $p$ and with zero orbital momentum; $M(\vec r)$ is the
amplitude of \pipi-system production at the relative distance $\vec r$.

Calculations of $M(\vec r)$ in the independent source model \cite{dirac}
allow to obtain \cite{af97} the following estimation:
\beq{e2}
\langle r \rangle_M = \frac{\displaystyle\int r M(r) d^3 r}
{\displaystyle\int M(r) d^3 r} \sim 5\div10\:\mathrm{fm} \ll r_B=1/\mu\alpha
=387 \:\mathrm{fm}\,,
\eeq
here $r_B$ is the pionum Bohr radius, $\mu=m_\pi/2$ is its reduced mass and
$\alpha$ is the fine-structure constant. Thus the most significant
contribution to the integrals in (\ref{e1}) comes from the short distance
where an influence of the strong interaction on the wave function can be
overruling \cite{kur98}.

To estimate the influence of the strong interaction on the considering ratio
(\ref{e1}) the results of paper \cite{kur98} on the pionium wave function
calculation in the perturbation theory can be applied:
\beq{e3}
\psi_{\mathrm{s}}(\vec r) =\psi_{\mathrm{s}}^{(C)}(\vec r) + \frac{\mu}{2\pi}
\int \frac{U(\vec{r'})}{|\vec{r}-\vec{r'}|} \;
\psi_{\mathrm{s}}^{(C)}(\vec{r'} )\:d^3 r' \,.
\eeq
Here $\psi_{\mathrm{s}}^{(C)}$ is the pure Coulomb wave function, i.e. the
wave function without account of the strong interaction, indices $s$
stand for $s$-wave functions of the discrete and continues spectra,
$U(\vec r)$ is the strong interaction potential of pions.

Then the ratio (\ref{e1}) can be rewritten as:
\beq{e4}
R(n\mathrm{s}/p\mathrm{s})=
\frac{\displaystyle\left|\int \widetilde{M}(\vec r)
\psi_{n\mathrm{s}}^{(C)}(\vec r)\,d^3r \right|^2}
{\displaystyle\left|\int \widetilde{M}(\vec r)
\psi_{p\mathrm{s}}^{(C)}(\vec r)\,d^3r \right|^2} \,,
\eeq
here
$$
\widetilde{M}(\vec r)=M(\vec r)+\frac{\mu U(\vec r)}{2\pi}
\int \frac{M(\vec{r'})}{|\vec{r}-\vec{r'}|}\;  d^3 r' \,.
$$
Obvious that the ``renormalized'' amplitude $\widetilde{M}(\vec r)$ is
short-range as initial $M(\vec r)$:
$$
\langle r \rangle_{\widetilde{M}} \sim \langle r \rangle_M  \ll r_B\,.
$$
It allows to use the power expansion of the wave functions
$\psi_{\mathrm{s}}^{(C)}$ at calculation of the integrals in (\ref{e4}).
\begin{eqnarray}
\psi_{n\mathrm{s}}^{(C)}(r)&=&\psi_{n\mathrm{s}}^{(C)}(0)
\left[ 1- \mu\alpha r +
\frac{1}{6}\left(2+\frac{1}{n^2}\right)(\mu\alpha r)^2 +
O\left((\mu\alpha r)^3\right)\right] \\
\psi_{p\mathrm{s}}^{(C)}(r)&=&\psi_{p\mathrm{s}}^{(C)}(0)
\left[ 1- \mu\alpha r +
\frac{1}{6}\left(2-\frac{p^2}{(\mu\alpha)^2}\right)(\mu\alpha r)^2 +
O\left((\mu\alpha r)^3\right)\right]
\end{eqnarray}

Finally the ratio (\ref{e1}) is written as:
\beq{e7}
R(n\mathrm{s}/p\mathrm{s})=
\frac{\displaystyle \left|\psi_{n\mathrm{s}}^{(C)}(0)\right|^2 }
{\displaystyle \left|\psi_{p\mathrm{s}}^{(C)}(0)\right|^2 }
\left(1+O(\mu^2\alpha^2\langle r^2 \rangle_{\widetilde{M}})
\right)\,,
\eeq
where the value of $O$ is of order $10^{-3}$. So it reproduces the formula
from the paper \cite{af97} which was obtained, basing on an erroneous results
of \cite{efim86,belk86}, for the pure Coulomb wave function of the
\pipi-atom. Thus in spite of the significant influence of the strong
interaction on the value of the pionium wave function at origin \cite{kur98}
the ratio of production rates of the pionium to Coulomb \pipi{} pairs is
expressed via the well-known Coulomb wave functions and in this way can be
calculated with required accuracy.

Authors would like to thank A.V.Tarasov for helpful discussions. This work
is partially support by RFBR grant 97--02--17612.

\end{document}